\documentclass{PoS}

\title{High-energy radiation from T Tauri stars}

\ShortTitle{High-energy radiation from T Tauri stars}

\author{{Gustavo E. Romero}\thanks{Member of CONICET.}\\
      Instituto Argentino de Radiastronom\'{\i}a
           (CCT La Plata, CONICET), C.C.5, 1894 Villa Elisa, Buenos Aires, Argentina\\
and\\
 Facultad de Ciencias Astron\'omicas y Geof\'{\i}sicas,
           Paseo del Bosque s/n, 1900 La Plata, Buenos Aires, Argentina\\
        E-mail: \email{romero@fcaglp.unlp.edu.ar}}

\author{{María V. del Valle}\\
Instituto Argentino de Radiastronom\'{\i}a (CCT La Plata, CONICET), C.C.5, 1894 Villa Elisa, Buenos Aires, Argentina\\
and\\
 Facultad de Ciencias Astron\'omicas y Geof\'{\i}sicas,
           Paseo del Bosque s/n, 1900 La Plata, Buenos Aires, Argentina\\
 E-mail: \email{maria@iar-conicet.gov.ar}}

\author{{Josep Mart\'{\i}}\\
Departamento de F\'{\i}sica (EPSJ), Universidad de Ja\'en, Campus Las Lagunillas s/n, 23071 Ja\'en, Spain\\
E-mail: \email{jmarti@ujaen.es}}

\author{{Pedro Luque-Escamilla}\\
Departamento de Ingenier\'{\i}a Mec\'anica y Minera, Escuela Polit\'ecnica Superior, Universidad de Ja\'en, 
Campus Las Lagunillas s/n, 23071 Ja\'en, Spain\\
E-mail: \email{peter@ujaen.es}}

\abstract{T Tauri stars are young, low mass, pre-main sequence stars surrounded by an accretion disk. These objects present strong magnetic activity and powerful magnetic reconnection events. Strong shocks are likely produced in the stellar magnetosphere by such events and charged particles accelerated up to relativistic energies.

We present a simple model for  the non-thermal radiation generated by high-energy particles in T Tauri stars. We discuss whether this emission is detectable at high energies with the available gamma-ray telescopes.}

\FullConference{25th Texas Symposium on Relativistic Astrophysics - TEXAS 2010\\
		December 06-10, 2010\\
		Heidelberg, Germany}

\begin{document}
\section{Introduction}
\begin{figure}
\begin{center}
\includegraphics[width=.6\linewidth]{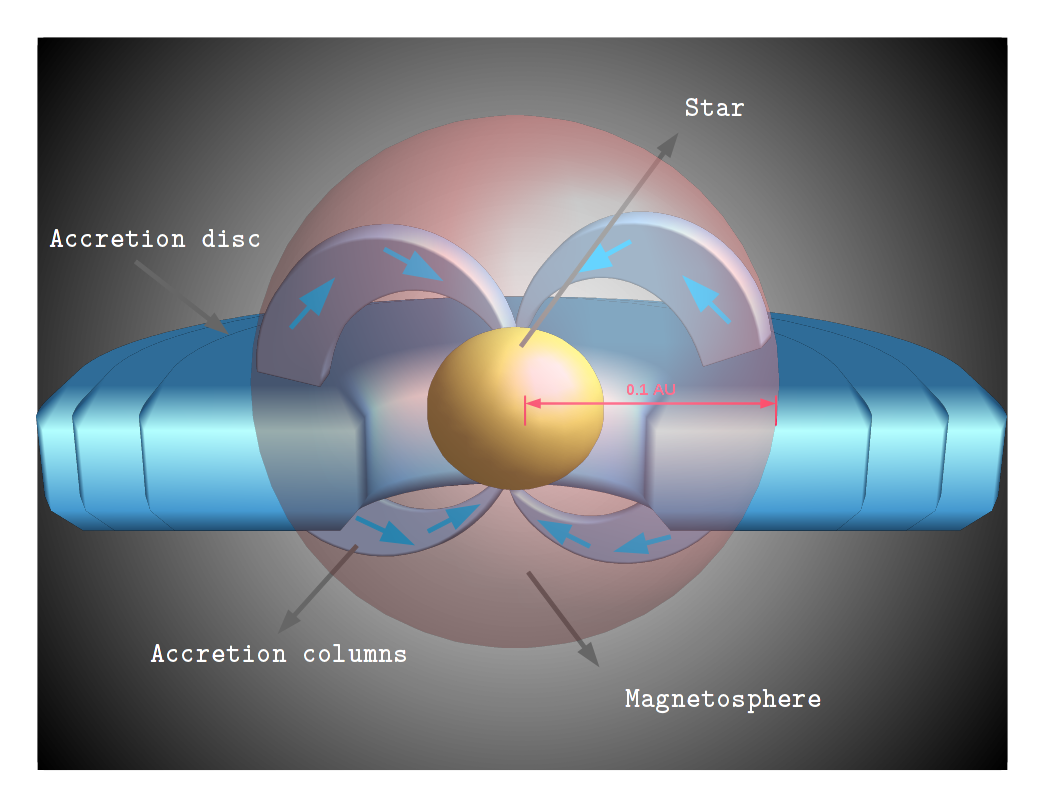}
\includegraphics[width=.38\linewidth]{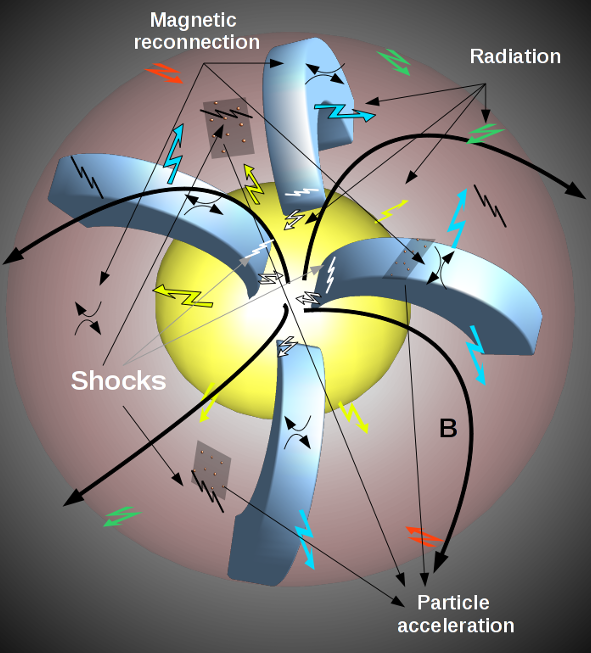}
\caption{T Tauri star scheme (not to scale)}
\label{fig:magnetosphere}
         \end{center}
\end{figure}

T Tauri stars are low-mass stars ($M$ $<$ 3 $M_{\odot}$) in their early stages of evolution. They are progenitors of solar-like stars. They  are surrounded by an accretion disk. The youngest objects drive bipolar outflows (see \cite{F}). The accretion is thought to be  magnetically confined and to proceed via infall along magnetic flux tubes threading the inner disk, leading to shocks and hot spots on the surface (e.g. \cite{ca}). Figure \ref{fig:magnetosphere} shows a simplified scheme of the physical scenario and a scheme of the physical processes which take place in the magnetosphere, viewed from the pole. 

Variable thermal keV X-ray emission is detected from T Tauri stars. 
This emission comes from a  high density plasma at a typical temperature of $\sim$ 10$^{7}$ K.
 X-ray flares are considered as upscaled versions of solar flares, related to  magnetic reconnection. 
 Several  works have been done on particle acceleration in  magnetic reconnection
 ( e.g. \cite{S}, \cite{D}). 
Non-thermal radio emission have being detected from these stars, therefore a population of non-thermal particles must exist in the system (e.g. \cite{Ph}, \cite{Jo}). 

 The most likely acceleration region is the
 magnetosphere, where violent magnetic reconnection episodes take
 place. Plasma collisions might produce strong shocks which can  accelerate
 particles up to relativistic energies through difussive Fermi-like mechanism. These
 particles can produce the non-thermal radio emission through synchrotron
losses. In this work we calculate the output of all losses of a lepto-hadronic particle population in the magnetosphere of T Tauri stars.

\section{The model}
We  consider that a power-law population of relativistic particles (electrons and protons) is injected in the magnetosphere. For simplicity we consider a spherical magnetosphere of radius $R_{\rm m}$. 
These particles interact with the magnetic field, with the various  radiation fields,
 and with the magnetosphere plasma.

\begin{figure}[h!]
           \includegraphics[width=0.5\linewidth,angle=270 ]{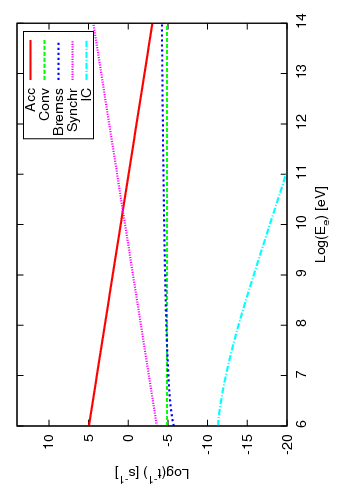}
 \includegraphics[width=0.5\linewidth,angle=270]{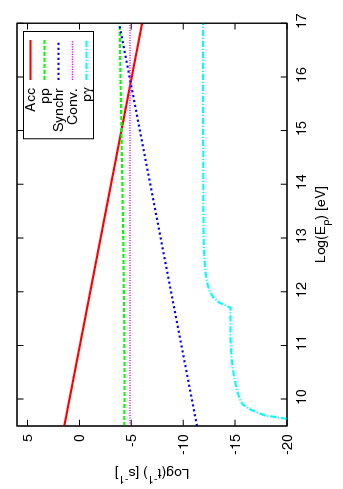}
                  \caption{Electron and proton losses and acceleration rates.}
\label{fig:Perdidas}
\end{figure}

 The values  adopted for the different parameters in our model are listed in  Table \ref{table}. 
We estimate the available power as $L = B^{2}/8\pi A c$ where $A$ is the magnetosphere area. We assume that 10\% of this power is release in the reconnection process. The efficiency of non-thermal acceleration is estimated in the Bohm limit  by $\eta\sim(v_{\rm s}/c)^{2}$, where $v_{\rm s}$ is the shock velocity. For solar-like values we obtain $\eta$ $\sim$ $10^{-4}$. 
\begin{table}
\caption[]{Parameters}
\begin{tabular}{lll}
\hline\noalign{\smallskip}
\multicolumn{2}{l}{Parameter } & value\\
\hline\noalign{\smallskip}
$R_{\rm m}$ & Magnetosphere radius & 0.1 AU  \\
$\eta$ & Acceleration efficiency& 10$^{-4}$\\
$a$ &Hadron-to-lepton energy ratio & 100 \\
$q_{\rm rel}$ & Content of relativistic particles & 10$^{-3}$   \\
$\alpha$ &Particle injection index & 2.2\\
$v_{\rm w}$ & Wind velocity & $2\times10^{8}$ cm s$^{-1}$   \\
$B$ &Magnetic field [G] & $10^{2}$  \\
$n$ & Maximum magnetospheric  density &  10$^{11}$  cm$^{-3}$ \\
$r_{\rm D}$ & Disk radius & 100 AU \\
$T_{\rm IR}$ & Disk temperature & 30  K\\
$T_{\star}$ & Star temperature & 4$\times 10^{3}$ K\\
$R_{\star}$ & Star radius & 2 $R_{\odot}$\\
$L_{\rm x}$ & X-ray luminosity & $10^{30}$ erg s$^{-1}$\\
\hline\\
\end{tabular}	
  \label{table}
\end{table}
The relativistic electrons lose energy mainly through synchrotron emission, inverse Compton (IC) scattering with the star radiation field and the X-ray radiation field, and through relativistic Bremsstrahlung with the magnetospheric plasma. The relativistic protons lose energy mainly through synchrotron emission and $pp$ inelastic collisions with ambient matter. Particles can escape from the acceleration region by wind convection.  The maximum energy for both types of  particles is obtained equating the cooling rates with the acceleration rate $t_{\rm acc}^{-1} = {\eta} e c B/E$. Figure \ref{fig:Perdidas} shows the different rates.

The particles steady state distribution is calculated using the standard  transport equation (see \cite{G}).  We also consider the population of secondary $e^{\pm}$ pairs injected by charge pion decay (e.g. \cite{O}).

We take into account four processes of interaction of the relativistic particles with the fields in the magnetosphere: 
synchrotron radiation of protons and electrons, $pp$ inelastic collisions, IC scattering and relativistic Bremsstrahlung (e.g. \cite{V}).
The particles will collide with the accretion plasma columns. These columns occupy  a non well-established volume of the magnetosphere. We adopt a small filling factor $f$ $\sim$ 10$^{-4}$.

\begin{figure}
  \begin{center}
  \includegraphics[width=0.4\linewidth,angle=270]{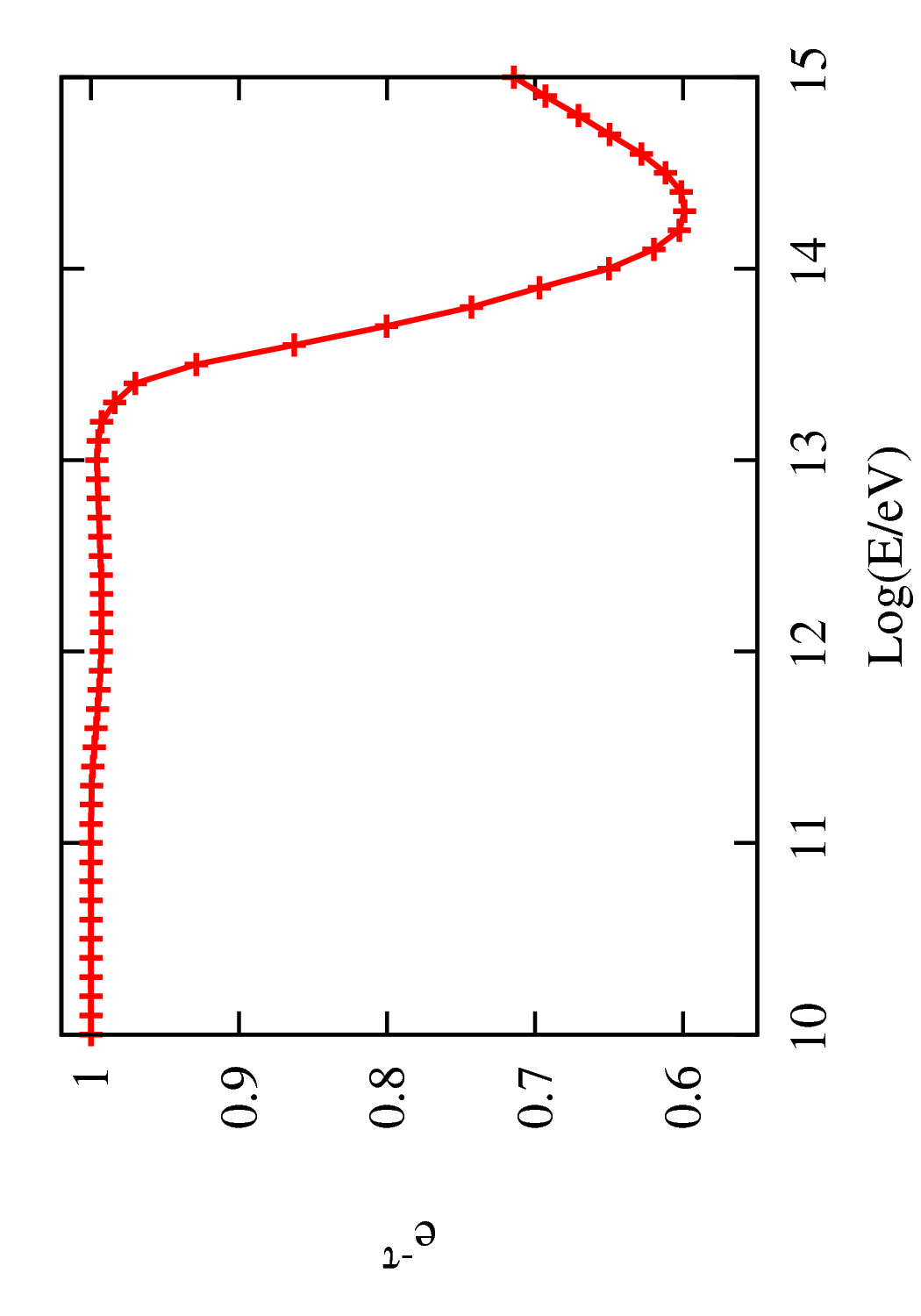}
\end{center}
\caption{Opacity curve.}
\label{fig:TAU}
\end{figure}

We also calculate the opacity from internal and external photon-photon absorption (e.g. \cite{Gu}). The external fields considered are the 
radiation fields from the disk and from the star,
and the X-ray thermal emission. The internal fields are the fields generated within the system by non-thermal processes. Fig. \ref{fig:TAU} shows the  opacity curve as a function of energy. The absorption is relevant for energies greater than 10 TeV.

\section{Results}
\begin{figure}[h!]
         \begin{center}
 \includegraphics[width=.6\linewidth,angle=270]{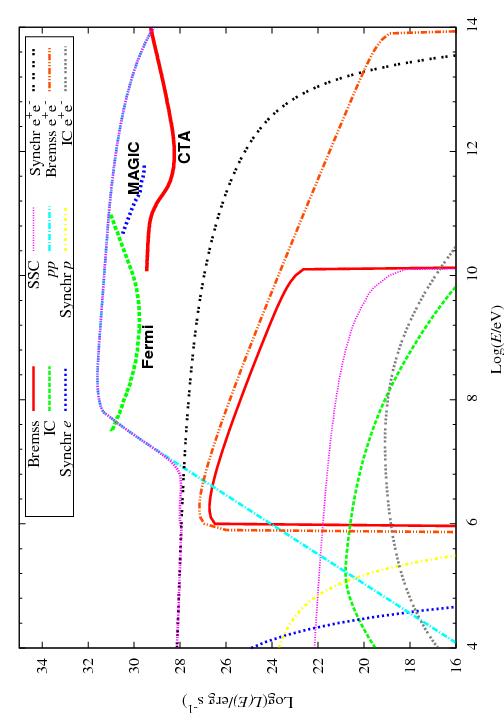}
                  \caption{Computed high-energy SED for a source at d $\sim$ 120 pc and the sensitivity curves for CTA, Fermi and MAGIC.}
\label{fig:SED}
         \end{center}
\end{figure}

Figure \ref{fig:SED} shows the computed spectral energy distribution (SED)  and the sensitivity curves from the gamma-ray detectors CTA, MAGIC, and Fermi. 
We consider a source at a distance $d$ $\sim$ 120 pc, similar to the distance of the nearest T Tauri stars in the ${\rho}$ Ophiuchus star forming region.

\section{Conclusions}
Some of the existing gamma-ray instruments
should be able of detecting T Tauri stars with the 
characteristic adopted in this work.

T Tauri stars might be the counterpart of some Fermi  galactic sources (e.g. del Valle et al. in prep.)
published in the first Fermi  LAT  catalog  (\cite{A}). Also  
Munar-Adrover, Paredes \& Romero 2011 (\cite{Mu}) have crossed the Fermi  catalog with catalogs of YSOs obtaining a set of candidates by spatial correlation. aproximately,  72\% of these candidates should be gamma-ray sources with a confidence level above 5$\sigma$. 
\section{Acknowledgements}
 This work was supported by PIP 0078 (CONICET), PICT 2007-00848 Prestamo BID (ANPCyT), by the Plan Andaluz de Investigacion, Desarrollo e Innovacion de la Junta de Andalucia as research group FQM-322 and excellence fund FQM-5418.


\begin{thebibliography}{99}
\bibitem{F}Feigelson, E.D. \& Montmerle, T. 1999, Annu. Rev. A\&A, 37, 363
\bibitem{ca}Calvet, N., \& Gullbring, E. 1998, ApJ, 509, 802
\bibitem{S}Schopper, R., Lesch, H., \& Birk, G.T. 1998, A\&A, 335, 26
\bibitem{D}de Gouveia Dal Pino, E.M., Piovezan, P.P., \& Kadowaki, L.H.S. 2010, A\&A, 518, id. A5
\bibitem{Ph}Phillips, R.B., Lonsdale, C.J. \& Feigelson, E.D. 1993, 403, L43
\bibitem{Jo} Johnson, K.J., Gaumo, R.A., Fey, A.L., de Vegt, C. \& Claussen, M.J. 2003, AJ, 125, 858
\bibitem{G}Ginzburg, V.L. \& Syrovatskii, S.I. 1964, The Origin of Cosmic Rays, Pergamon Press, Oxford
\bibitem{O}Orellana, M. et al. 2007, A\&A, 476, 9
\bibitem{V}Vila, G.S. \& Aharonian, F.A. 2009, in Compact Objects and their Emission, Romero, G.E. \& Benaglia, P. (eds.), Paideia, La Plata, P. 1
\bibitem{Gu}Gould, R.J. \& Schr\'ereder, G.P. 1967, Pys. Rev., 155, 1404
\bibitem{A}Abdo, A.A. et al. Fermi LAT coll 2010, ApJ Supp., 188, 405
\bibitem{Mu}Munar-Adrover, P., Paredes, J.M., \& Romero, G.E 2011, 
proceedings of the  IAU Symposium 275, 275, 406

\end{thebibliography}
\end{document}